\documentclass[runningheads]{llncs}
\usepackage{graphicx}
\usepackage{amssymb}
\usepackage{hyperref}
\usepackage{cleveref}
\usepackage{tabularx}
\usepackage{graphicx}
\usepackage{adjustbox}
\usepackage{array}
\authorrunning{Short author list}% Part of LEFT running header
\titlerunning{EDoS Mitigation Techniques in Cloud Computing Environment}% Part of RIGHT running header

\begin{document}
\title{Comparative Analysis of State-of-the-Art EDoS Mitigation Techniques in Cloud Computing Environment}
\author{Parminder Singh\inst{1} \and Shafiq Ul Rehman\inst{2} \and Selvakumar Manickam\inst{1}}
\authorrunning{P. Singh et al.}
\institute{National Advanced IPv6 Centre (Nav6), Universiti Sains Malaysia (USM), Penang, Malaysia \and
ST Engineering Electronics - SUTD Cyber Security Laboratory, Singapore University of Technology and Design (SUTD), 8 Somapah Road, 487372, Singapore 
\email{shafiq\_rehman@sutd.edu.sg} }
\maketitle              % typeset the header of the contribution
\begin{abstract}
A new variant of the DDoS attack, called Economic Denial of Sustainability (EDoS) attack has emerged. Since the cloud service is based on the “pay-per-use” model, the EDoS attack endeavors to scale up the resource usage over time to the point the purveyor of the server is financially incapable of sustaining the service due to the incurred unaffordable usage charges. The implication of the EDoS attack is a major security implication as more elastic cloud services are being deployed. Existing techniques to detect and mitigate such attacks are either have low accuracy or ineffective and, in some cases, aggravate the attack even further. Therefore, an enhanced EDoS Mitigation Mechanism (EMM), is proposed to address these shortcomings using OpenFlow and statistical techniques, i.e. Hellinger Distance and Entropy. The experiments clearly depicted that EMM is able to detect and mitigate EDoS attacks with high accuracy and it is effective in terms of resource utilization compared to existing mitigation techniques. Thus, it can be deployed in the cloud environment without the need for additional resource requirements.

\end{abstract}
\section{Introduction}
Cloud Computing \cite{velte2010cloud,kuyoro2011cloud}, is the next revolution in the Information and Communication Technology (ICT) arena \cite{adamov2009truth}. It is a model in which computing is delivered as a commoditized service like electricity, water, and telecommunication. Cloud computing (CC) provides software, platform, infrastructure, and other hybrid models, which are delivered as subscription-based services in which customers pay based on usage \cite{bhardwaj2010cloud}. Nevertheless, security is one of the main factors that inhibit the proliferation of CC \cite{zissis2012addressing,popovic2010cloud}.

Economic Denial of Sustainability (EDoS) is a new breed of security and economical threats to the CC paradigm \cite{singh2014survey}. Unlike the traditional Distributed Denial of Service (DDoS) \cite{hoff2008cloud} which brings down a service by exhausting the resources of a targeted server in traditional setup, EDoS takes advantage of the elasticity feature provided as a cloud service \cite{hoff2008cloud,swami2019software}. This causes the resources to dynamically scale to meet the demand (because of EDoS attack) resulting in a hefty bill for the customer. EDoS attack is considered as one of the concerns that has stalled many organizations from adopting or migrating their operations and/or data to the cloud \cite{singh2014survey}. This is because an EDoS attack targets the financial component of the service provider and customers; which has high visibility on the impact of such attacks.

In this research, the detection and mitigation accuracy of EMM is assessed and compared against the state-of-the-art techniques such as EDoS Shield and with other techniques in terms of various quality facets. Moreover, the mitigation performance of the EMM is also assessed against HTTP and UDP attacks. In both cases, the results demonstrate that the EMM improves the mitigation capabilities of cloud systems. In all scenarios, EMM can detect HTTP and UDP attack resulting in improved mitigation accuracy. The assessment result demonstrated that EMM achieved its objectives.

Rest of the paper is structured as follows: Section \ref{ssec:sec2} describes some of the existing state-of-the-art EDoS mitigation techniques, their features and limitations. The proposed EMM technique is presented in Section \ref{ssec:sec3}. The experimental setup is depicted in Section \ref{ssec:sec4}. Performance evaluation of the proposed scheme and its comparative analysis with existing EDoS techniques are presented in Section \ref{ssec:sec5}. And finally, the conclusion and future works are outlined in Section \ref{ssec:sec6}.

\section {Existing EDoS Mitigation Techniques }
\label{ssec:sec2}
Most of the current literature available addresses mainly on DDoS protection emphasizing on techniques for preventing of apparently malicious traffic at the network or application layer \cite{swami2019software,chaudhary2018survey,joshi2012securing,chapade2013securing}. There is limited number of literatures that are available to provide deployable solutions specifically for mitigating EDoS attacks in Cloud Computing environment. Most of the researchers in the field of CC and network monitoring are relying on the predefined threshold and on entropy techniques to detect anomalies in network traffic. Some of the well-known EDoS mitigation techniques are as follows:

In 2009 self-verifying proof of work (sPoW) was proposed by Khor and Nakao \cite{khor2009spow}. The authors attempted to distinguish the EDoS attack, On-demand network filter and prioritize the legitimate traffic from network level DDoS traffic. In order to do so, sPoW does two things: First, it does packet pattern matching to distinguish and filter the DDoS network traffic. Second, it allows the legitimate traffic to go through the network. This mechanism does packet filtering based on cryptography puzzles methodology. By applying such algorithm, it allows legitimate traffic while discards the malicious network level DDoS traffic. However, it possesses certain drawbacks. Firstly, asymmetric computational power consumption for the clients. Solving computational puzzles require more CPU power and suitable only to faster CPUs. Therefore, mobile devices with less processing power will not be able to re-solve the puzzles, thus unable to access the cloud resources \cite{green2011reconstructing}. Secondly, the Server must create separate channels to address each request. In case of large number of incoming requests, server will generate number of puzzles which leads to puzzle accumulation attack if puzzles do not resolve in time.

Later in 2011, Sqalli et al. proposed another mitigation technique called EDoS-Shield \cite{sqalli2011edos}. This technique distinguishes the legitimate and malicious requests by checking the user’s presence at client-side machine. EDoS-Shield architecture consists of Virtual Firewall (VF) and Verifier Nodes that functions in tandem to execute the EDoS mitigation tasks. The incoming requests are filtered by firewall based on white and black-lists. When client makes access request, the verifier node verifies it through a Turing test. In case client passes that test, its IP address will be maintained in white-list and the subsequent requests received from the same client are forwarded directly to the cloud scheduler, approving resource allocations. Whereas, if a client fails that Turing test, its IP address will be included in the black-list and the subsequent requests from the same client will be dropped down by the firewall itself. Nevertheless, the proposed technique has certain constraints. First, it is vulnerable to spoofing attack. By performing IP address spoofing, attacker can use IP address belonging to the white-list of the verifier node to carry out an EDoS attack that would remain undetected. Second, due to its designed mechanism it can possess higher false positive rate by blocking of many IP addresses belonging to the legitimate users, as the two lists are not updated in a timely and accurate manner. 

In-cloud scrubber service is another mechanism proposed to mitigate the EDoS attack \cite{kumar2012mitigating}. This mechanism uses on-demand web service (Scrubber Service) to address such attack by using crypto puzzle approach to validate the legitimate users request \cite{khor2009spow}. This mechanism offers two modes namely; normal mode and suspected mode. Based on the requirement service providers can choose the one accordingly. For instance, during normal scenario when the web server is perceived normally it runs in normal mode. Whenever service provider observes the web server resource exhaustion beyond an acceptable limit and high bandwidth utilization, this could be considered as attempt of EDoS attack. Thus, service provider enables its suspected mode and an On-demand call is directed to the Scrubber service to generate and verifies hard puzzle. But there are certain limitations with this mechanism as follows: it can only detect and mitigate HTTP attacks, also it depends on third party application for authentication purpose.

In 2013, Mudassar proposed an EDoS mitigation framework known as EDoS Armor \cite{masood2013edos} specifically meant for E-Commerce applications. It uses two-step approach, comprises of admission control and congestion control. At first, when user initiates a session, the server sends a challenge to the user, it could be an image or a cryptographic puzzle to be resolved, in case user resolves the challenge, the request is being directed to admission control. Otherwise, the session of the user gets dropped and the number of connections to the server are limited for the user. This technique applies port hiding approach to restrict the users, as attack cannot be initiated in the absence of a valid port number. Next, the behavior of the user browsing is consistently being traced. whenever an abnormal behaviour is detected, service priority is being set for that particular user based on his previous records. Thus, in this manner, EDoS Armor can mitigate EDoS application. However, in E-commerce applications new users may find it complicated system due to its complexity in design. Therefore, in practice, the implementation of this method seems to be doubtful.

Recently, Chowdhury et al. proposed a new approach known as EDoS Eye to counter the EDoS attack \cite{chowdhury2017edos}. This model applies game theory approach to mitigate the EDoS traffic. Authors claim to develop a Game Based Decision Module (GBDM) that can get threshold values to restrict the incoming traffic. Similarly, another technique proposed by Shawahna et al. is known as EDoS Attack Defense Shell (EDoS-ADS) \cite{shawahna2018edos}. According to the authors, this technique is applicable to Network Address Translation (NAT) based networks where it gets triggered if it sniffs any incoming abnormal traffic. For instance, in an attack scenario, it triggers a checking component to differentiate between legitimate users and attackers. However, both techniques were evaluated in simulation environments that have their own limitations as well as not closely representing the real-world scenario. 

In the context of this work, a summary of previous efforts and research has been done in the area of network worm, network scanning and signature automation approaches. The main observations that can be derived from the literature review are: 
\begin{itemize}
    \item It is inevitable that CC is replacing traditional computing, and this has consequently brought along security challenges unique to cloud infrastructures. 
    \item EDoS, a variant of DDoS attack, specific to subscription-based CC services, is on the rise. Such an attack affects the financial aspect of hosting a service on a CC environment. 
    \item 	Most of the existing techniques to addressing the issue of EDoS attacks are mainly Application-based solutions. 
    \item  An appropriate testbed that can enable real-time and hi-speed analysis of a net-work is lacking. 
    \item A more robust approach is required to ensure EDoS can be mitigated effectively. 
\end{itemize}

\section{Our Proposed Technique}
\label{ssec:sec3}
\begin{figure}[ht!] 
\includegraphics[width=\textwidth]{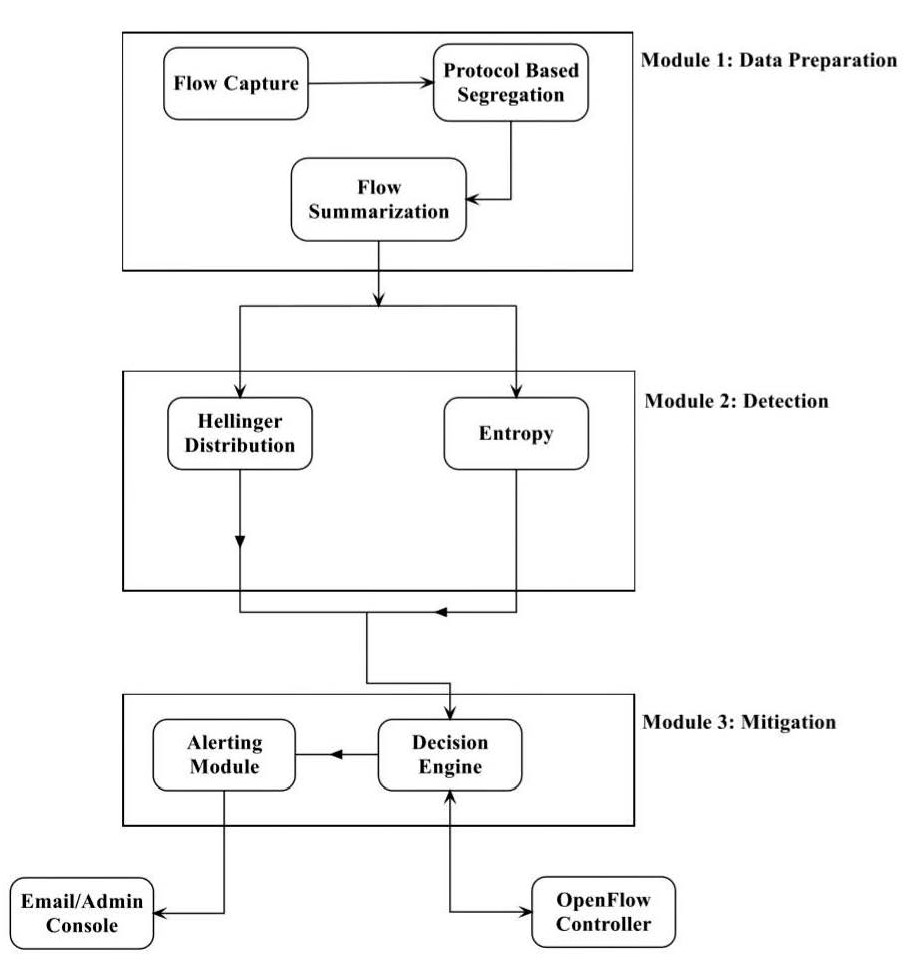}
\centering
\caption{Proposed EMM technique.} 
\label{fig:fig1}
\end{figure}

In this section, we describe our proposed mitigation mechanism for Cloud Compu-ting (CC) environments known as EDoS Mitigation Mechanism (EMM) \cite{bawa2017enhanced}. The aim of this mechanism is to be effective enough to mitigate different type of EDoS attacks while consume less resources. It consists of three major modules namely Data preparation, Detection, and Mitigation, which work in conjunction to achieve this objective. Fig.\ref{fig:fig1}, depicts the architecture of the proposed EMM technique.

In first phase, the network traffic is sampled using sFlow agents and sent for analysis to sFlow collector. Network statistics are generated using the inputs received by the sFlow collector and traffic is then segregated based on source IP, source port, destination IP, destination port, and counter. In the second phase, network statistics are compared against the threshold value defined using HD and classified for suspicious behaviours based on an entropy analysis. In the final phase, Open Flow (OF) controllers drop the network traffic of suspicious source IP addresses by updating switching rules and continue monitoring the network for anomalies. The functionality of these EMM components are described as:

\subsection{Data preparation}
In this module, data is being collected during the flow-based monitoring \cite{shalimov2013advanced,hulboj2007packet}. sFlow agent is used to collect the flow of network traffic and passed to sFlow collector for information extraction. EMM leverages sFlow packet sampling technique to monitor traffic in real-time, Packet sampling decouple the flow collection process form the forwarding plane and provides all flow related statistical information. This method provided efficient and aggregated packet forwarding, eliminating the specific flow entries requirement of native OF approach and overcome flow-table size limitations by reducing the number of flow entries in OF switches. EMM uses a simplified flow collection algorithm to minimize the system resource requirements and provides adequate information for a reliable attack detection process. 

Based on the protocol type such as TCP, UDP or ICMP, the collected data is processed and segregated before being summarized and later passed to detection module. The 6 tuple information like switch ID, source IP, destination IP, source port, destination port and counter from the datagram are being extracted. This information is later utilized by detection module to set a dynamic threshold using Hellinger distance and entropy based flood detection for alerting and mitigation of attack. 

\subsection{Detection module} 
The collected datagram packets are being analyzed to extract the information such as source and destination IP, port number, and number of packets per second. By using Hellinger’s distribution \cite{sengar2008detecting} and entropy methods \cite{shannon1948note}, this module defines the dynamic threshold settings as well as performs anomaly detection tasks respectively.
\subsubsection{Threshold Detection:}
Hellinger’s distribution (HD) is used to measure the distance between two probability distribution. To compute HD, assume two distributions on same sample space are present, namely:\emph{ P:(p1, p2, ..., pn)} and \emph{Q:(q1, q2, ..., qn)}. HD between two distributions can then be defined as: 
\begin{equation}
   H^2(P, Q) = \frac {1} {2} \sum {_i^n} = 1 (\sqrt{Pi} - \sqrt{Qi})^2
    \end{equation}
    
HD value will vary between 0 and 1, where 0 represent identical distribution and 1 represents different probability distribution. Low probability among two distribution implies no significant deviation and abrupt elevated HD simply indicates the anomaly or attack in the network. To indicate the anomaly in the network, a detection threshold is required. To obtain a dynamic threshold that allows the proposed mechanism to be used in any kind of network environment, EMM relies on HD probability distribution method \cite{bawa2017enhanced}.

We adopted entropy method \cite{shannon1948note} in our proposed EMM technique as the anomaly detection algorithm. The opted method not only effectively classifies attack patterns, but also distinguishes the attackers and the victims. Once network anomaly is detected, this method examines and correlates definite network metrics identifying the attack and revealing all related information to the attack mitigation module.

\subsection{Mitigation module}
This is the last module; it is accountable for alert generation and mitigation of attacks in the network. By updating the rules on the network devices, it takes appropriate action. Decision engine analyses the incoming traffic (received from mitigation module) against a set of rules. Based on that analyses decision such as whether to block the network traffic originating from an IP address for specific period or not can be made.

The processed information from detection module is passed to decision engine where a dynamic threshold is applied to detect the EDoS attack and entropy analysis is applied to verify the existence of attack. Both feedbacks from detection module is correlated with the network traffic statistics from the OpenFlow network Switch. Based on the correlation, the decision engine (DE) in tandem with the mitigation engine make decision to either drop the packet on network perimeter or report the anomaly to the network/client administrator.

Alerting engine allows EMM to generate alert and update the client’s cloud administrators via periodic email updates. In case of an anomaly detected in the network, this engine sends network updates using its periodic update cycle as configured by client. Whereas, in case of an attack, it immediately sends a message or an E-mail alert to the client as well as CSP’s Security Operation Centre (SOC) for notifying the ongoing attack. 

\section{Experimental Setup}
\label{ssec:sec4}
Before going to the results of the evaluation, it is imperative to understand the design of the experiments upfront. They are detailed as follows:

\subsection{Cloud Testbed Setup Environment}

The performance of any of such mechanism is best evaluated on an existing cloud computing environment. However, access to such an environment is often not possible due to the usage of propriety tools or service provider organization’s policies. Although there are solutions using simulation tools such as Mininet \cite{de2014using}, only limited evaluation can be performed. For this purpose, a standard benchmark for cloud computing environment is developed as a testbed using OpenStack cloud computing environment with OpenDay Light (ODL) as Open Flow Controller (OFC) \cite{shalimov2013advanced}. This allows the possibility to create network topologies and to carry out various tests.

An OpenFlow controller is an application, which controls the network flow in SDN. Generally, SDN uses OpenFlow protocol to manage the network. OF acts as an operating system for virtual networking. Communication between devices and applications must pass through the controller and controller will update flow-tables on switches to send data packets to application. The controller that manages network devices (Specification-Version 2013) and decides on the best route to route traffic to application uses the OpenFlow protocol. Instead of hardware firmware, network control plane is used so that network can be managed more dynamically and with increased precisions. 

OpenDaylight (ODL) controller is the most recent addition to OpenFlow controllers and is written in Java. It is meant to be a common platform for all SDN users. Recently, ODL pronounced its second release, Helium. ODL support most of the operating systems, i.e., Linux, Mac and Windows, and it has the feature of topology discovery. ODL uses karaf framework, which makes it modular, besides enabling plug-in of various application modules developed in Java.
The developed testbed enables the opportunity for one to utilize a modern platform to design the topology with any number of virtual machines, OpenFlow controllers, virtual routers, and firewalls. With the implemented testbed, virtual machines can be launched with various web applications (web server). The testbed also consists of target machine, which primarily experience the large volume of attach/user traffic.

At first instance, all new traffic will pass through OFC, which will update all the network switches to perform basic forwarding functions. However, all switches and components are configured with sFlow agents to periodically update the monitoring agent with the current network statistics. Once traffic exceeds a defined threshold, the decision engine will update the OFC about the network condition and in return, OFC updates the flow table of switches to instruct the devices to drop the traffic from a specific attacker host. The rest of the traffic from other nodes will not be affected by the new flow update. The design of this testbed allows efficient monitoring process without introducing latencies or overhead to the underlying cloud computing network.

The specification of the utilized hardware (servers) is provided as below and in Table \ref{table:tab1}, Blade Chassis M100OC with Power Edge M 610 Server: 
\begin{itemize}
\item M610, Dual Quad Core Processor / 16 GB / 1 TB HDD.
\item 720xd, Dual Hex Core Processor/ 64GB/ 4 TB HDD.
\end{itemize}

\begin{table}[ht!]
\caption{Hardware Specifications} % title of Table
\centering % used for centering table
\begin{adjustbox}{width=75ex}
\begin{tabular}{c| c| c | c} % centered columns (4 columns)
\hline\hline %inserts double horizontal lines
No. of Machines & Operating Systems &	Hardware Details &	Purpose \\ [0.5ex] % inserts table
%heading
\hline % inserts single horizontal line
1 &	Ubuntu14.04 x64 &1TB HDD, 16GB RAM &	Controller \\  [1ex] % [1ex] adds vertical space % inserting body of the table
1 &	Ubuntu14.04 x64 &	1TB HDD, 16GB RAM &	Network \\ [1ex] % [1ex] adds vertical space
1 &	Ubuntu14.04 x64	& 1TB HDD, 16GB RAM	& OpenFlow Controller \\ [1ex] % [1ex] adds vertical space
1 &	Ubuntu14.04 x64 &	1TB HDD, 16GB RAM &	Compute-1 \\ [1ex] % [1ex] adds vertical space
1 &	Ubuntu14.04 x64 &	1TB HDD, 16GB RAM &	Compute-1 \\ [1ex] % [1ex] adds vertical space
1 &	Ubuntu14.04 x64 &	1TB HDD, 16GB RAM &	Compute-1 \\ [1ex] % [1ex] adds vertical space
1 &	VMware ESXi6 &	4TB HDD, 64GB RAM &	Various VM \\
\hline %inserts single line
\end{tabular}
\end{adjustbox}
\label{table:tab1} % is used to refer this table in the text
\end{table}

\section{Performance Evaluation}
\label{ssec:sec5}

All conducted experiments were carried out for a duration of 5 minutes. The duration of 5 minutes was used because the auto scaling timers used for the upper threshold is assumed to be duration of 5 minutes \cite{baig2016controlled}. Specifically, the evaluation was performed to answer the following research question:

a) What is the influence of network traffic from different communication protocols, i.e., HTTP, (TCP) and UDP in the effectiveness of EMM in a cloud computing environment?

b) How well does EMM perform in comparison with other state-of-the-art approach?

The network topology for all the evaluation consists of five nodes that generate (pushes) normal traffic to the target machine and one host will be dedicated for inducing and generating attack traffic to the target machine. Victim’s network bandwidth is fixed at 10Mbps so that any variation in network can be easily measured. 

\subsection{ Evaluation Metrics}

The core idea behind the proposal of a mechanism i.e. EMM is to quickly detect and mitigate an ongoing EDoS attack as soon as possible to reduce the incurred damage to the cloud service/user. As such, the experiments that were conducted in evaluating EMM uses the standard metrics which measures the consumed resources, such as CPU, Memory, and Network bandwidth utilization rates.

The evaluation metrics are to be interpreted as following: the closer the resource consumption pattern to the baseline (or normal traffic), the better the performance of the evaluated mechanism. This is also applicable to the costs incurred by the cloud user; whereby, the lower the costs incurred, the better the performance of the mechanism in place.

\emph{Use Case 1: Influence of TCP (HTTP) Flooding Attack as an EDoS Attack.}

The first set of experiments were conducted to investigate the influence of HTTP traffic flooding attack as an EDoS attack. HTTP flooding attack is the most common application layer attack (OSI Layer 7). This attack is volumetric in nature and uses legitimate protocol’s GET or POST requests to retrieve information from the URL data. 
In the first 

In the first experiment, the victim machine was flooded with HTTP GET requests using the BoNeSi botnet DDoS HTTP flood simulator. The network and computer’s network utilization are constantly being monitored via the monitoring tools. 

At the same time, network traffic of random HTTP GET requests are generated using scripts that are running on five participating nodes to generate normal traffic to the webserver. The command line for the script is executed as shown below: 
\begin{verbatim}
root@ubuntu:\$perl httpflooder.pl -a GF –h 192.168.200.53 –t 400
\end{verbatim}

\emph{Use Case 2: Influence of UDP Flooding Attack as an EDoS Attack.}
%\textit{Use Case 2: Influence of UDP Flooding Attack as an EDoS Attack.}

A UDP flooding attack is another most common volumetric attack in the network. Attackers craft large-sized UDP packets and direct them to the victim. As the UDP protocol is “connectionless”, it can be exploited to launch quick and massive attacks as it does not require any handshakes as it is with TCP sessions.

In fact, UDP attacks on NTP service have been observed to have reached a peak of 500Gbps \cite{hulboj2007packet}. DNS amplification attack is another example of UDP attack also known as the alphabet soup attack. As there is no specific packet format defined for UDP, attackers can craft large packets, fill it with junk text or numbers (“alphabet soup”), and redirect them to the victim of their choice. The victims usually have to receive the packets and analyze them manually one by one before discarding useless packets.

Such a UDP attack has been used to evaluate the effectiveness of EMM to mitigate such attacks. For that, IPerf is used to generate UDP streams of 20Mb of random size packets to the victim IP as the attack traffic. The attack is repeated with an interval of 400 sec using command as: 

\begin{verbatim}
root@ubuntu:$ iperf -c 192.168.200.53 –v –b 20M –t 400
\end{verbatim}

In addition to the attack traffic, normal UDP traffic is also generated using 10Mb as bandwidth for a 300 sec interval using IPerf command line as shown below: 
\begin{verbatim}
root@ubuntu:$ iperf –c 192.168.200.53 –u –b 10M –t 300
\end{verbatim}

\subsection{Comparative Analysis}
To understand the influence of the different types of protocols on the performance of EMM, an experiment on comparing the effectiveness of EMM against the state-of-the-art technique was conducted. The EDoS Shield framework has been chosen and implemented in the testbed to be compared with EMM because of its efficiency in HTTP GET flooding attack mitigation as mentioned in Section \ref{ssec:sec2}. The proposed EMM and EDoS shield are compared using the evaluation setup that was used for EDoS shield \cite{sqalli2011edos}:

\begin{itemize}
    \item Normal network traffic rate 48Kpps.
    \item Attack traffic rate of 184Kpps.
    \item EMM is configured to block attacker host for an hour (similar to EDoS shield).
\end{itemize}

This comparative study conducted in two scenarios: (a) attack originated from random hosts, and (b) attack originated from a white-listed hosts. The results of the comparative evaluation are presented as follows:

\subsubsection{Attack Originated from a Random Host (non-whitelist):} 

In this experiment, normal traffic rate of 48Kpps is generated from one of the hosts using the httpflooder.pl script and the attack traffic of 184Kpps is generated from five host using BoNeSi tool for a duration of one hour.
In EDoS-Shield, as a legitimate host try to access the services on cloud, it gets verified by the firewall and verifier node using the GTT test. Once verified, IP of the host gets added to the whitelist of the firewall and gets subsequent access to the cloud resources. When an (automated) attacker host access the service, it would fail to respond to GTT test and gets added to the blacklist of the firewall which allows only legitimate user to access the cloud resources.

Meanwhile, in EMM, legitimate hosts trying to access the service on cloud will be serviced by the cloud resources. However, when an attacker requests access to the resources, it too is allowed as long the requests are within the allowed dynamic threshold. Any attempt of abusing the resource consumption would lead the attacker’s IP to be blocked for a pre-defined duration, i.e., an hour.

\begin{figure}[ht!]
\includegraphics[width=\textwidth]{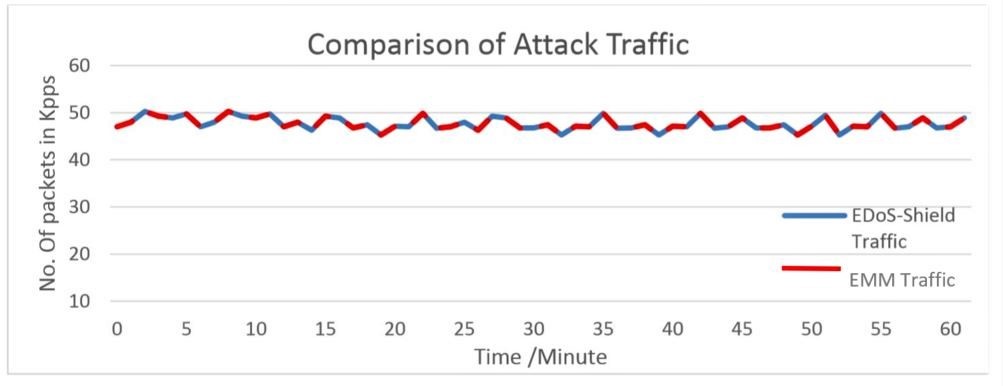}
\centering
\caption{Network utilization comparison between EMM and EDoS-Sheild in new attack scenario.} \label{fig:fig2}
\end{figure}

\begin{figure}[ht!]
\includegraphics[width=\textwidth]{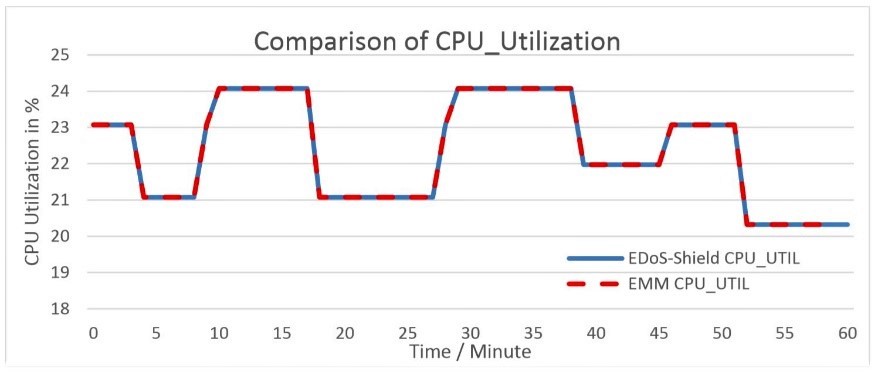} 
\centering
\caption{CPU utilization comparison between EMM and EDoS-Sheild in new attack scenario.} \label{fig:fig3}
\end{figure}

\begin{figure}[ht!]
\includegraphics[width=\textwidth]{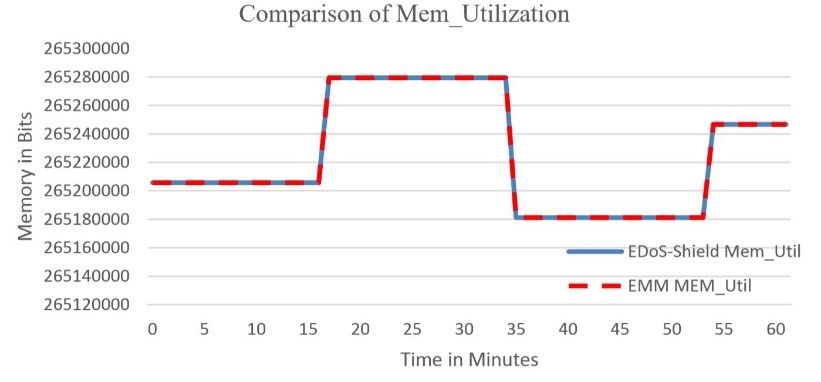}
\centering
\caption{Memory utilization comparison between EMM and EDoS-Sheild in new attack scenario.} \label{fig:fig5}
\end{figure}

The analysis of the data collected from the comparative experiments are presented in Fig. \ref{fig:fig2}, \ref{fig:fig3}, and \ref{fig:fig5}. In general, the performance of both EMM and EDoS Shield is roughly the same across the different measurements.
Fig. \ref{fig:fig2} depicts the network utilization pattern (in terms of Kpps) in dependence of time. EMM and EDoS-shield performs similar across the experiments with very minor fluctuations that are negligible.

Fig. \ref{fig:fig3} shows the CPU utilization rate between both compared mechanisms. Both mechanisms utilized pretty much the same amount of CPU utilization to perform in mitigating the EDoS attack.

Fig. \ref{fig:fig5} illustrates the memory utilization comparison between the two mitigation mechanisms. Throughout the evaluation, the memory utilization between the two mechanisms remained the same.

\subsubsection{Attack Originated from a White-listed Host:}
In this scenario, normal traffic of 48Kpps is generated from one host using the httpflooder.pl script and the attack traffic of 184Kpps is generated from five hosts using BoNeSi for the duration of one hour.

In EDoS-Shield, as the attacker host and legitimate host are already in the white-list, hosts trying to access the services on cloud will be granted directly. Hence, this would be a scenario whereby the network/victim do not have any mitigation mechanism in place.
Whereas in EMM, as legitimate hosts try to access the service on cloud, they will get serviced with the requested cloud resources. This also means that the attackers would be allocated relevant resources as long as the dynamic threshold level is not breached. If the dynamic threshold is exceeded, all network traffic with the attacker’s source IP address will be blocked for an hour.

\begin{figure}[ht!]
\includegraphics[width=\textwidth]{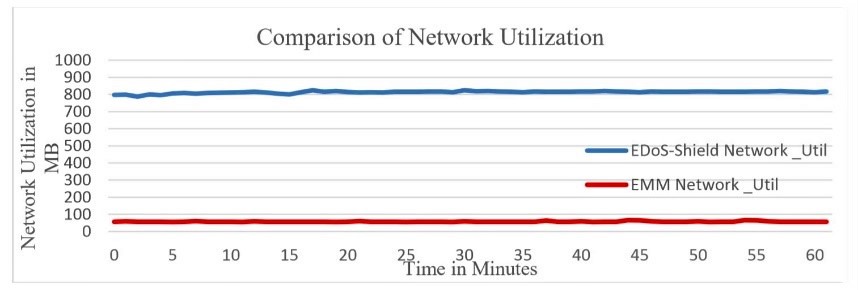}
\centering
\caption{Network utilization comparison between EMM and EDoS-Sheild in white-listed attack scenario.} \label{fig:fig6}
\end{figure}

Fig. \ref{fig:fig6} depicts the measurement of the network utilization during the attack with each mechanism in place. When EDoS shield is in place, the attack traffic surges up to 800Mb of network utilization. Meanwhile, for EMM, the utilization rate remains below 100Mb throughout the experiment. Another observation that can be inferred from the results of the analysis is: EMM is able to protect the cloud users by conserving up to 700Mb of the network bandwidth utilization.

\begin{figure}[ht!]
\includegraphics[width=\textwidth]{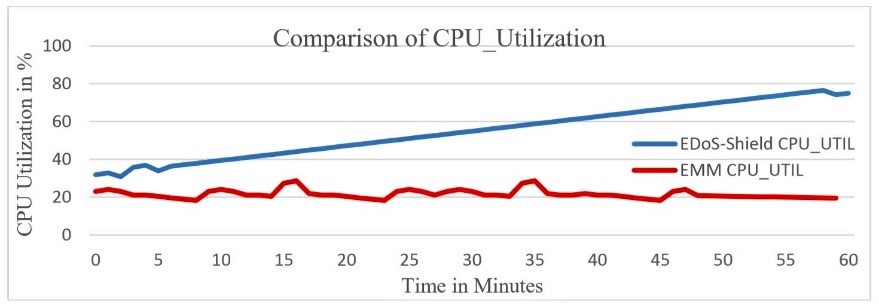}
\centering
\caption{CPU utilization comparison between EMM and EDoS-Sheild in white-listed attack scenario. } \label{fig:fig7}
\end{figure}

Fig. \ref{fig:fig7} shows the measurement of the CPU utilization during the attack with each mechanism in place. When EDoS shield is in place, the attack traffic surges up to 80\% of utilization rate. Meanwhile, for EMM, the utilization rate remains below 30\% throughout the experiment. Another observation that can be inferred from the results of the analysis is: EMM is able to protect the cloud users by conserving up to 45\% of the CPU utilization rate.

\begin{figure}[ht!]
\includegraphics[width=\textwidth]{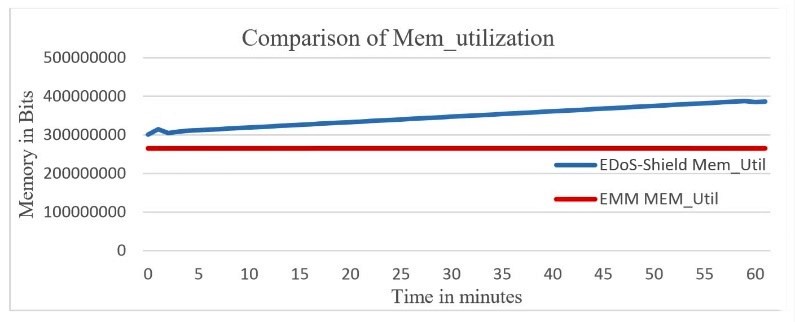}
\centering
\caption{Memory utilization comparison between EMM and EDoS-Sheild in white-listed attack scenario.} \label{fig:fig8}
\end{figure}

Fig. \ref{fig:fig8} represents the memory utilization of the victim machine under attack condition with the two mitigation mechanisms in place. Since EDoS-shield is not able to filter the traffic origination from white-listed IP addresses, the memory consumption has increased slightly more than 50\% from that of the consumption for EMM. In the case of cloud computing services, when the initially allocated resources are depleted or almost exhausted, additional VMs are allocated. However, these translates as additional costs for the providers of the service, i.e., cloud users. This is further exploited to cause EDoS attack on them by the attacker.

To summarize the experiment in comparing the performance and features of EMM and EDoS Shield:

\begin{itemize}
\item EDoS Shield utilizes a firewall, black-list, and white-list modules. If an IP from white-list initiate an attack, this framework does not provide any kind of protection. Hence, this is a major drawback of this mechanism. Meanwhile, EMM uses network traffic only. When the dynamic threshold is exceeded, all packets from the attacker’s source IP will be blocked for a pre-defined interval. Hence, this mechanism is not susceptible to any direct attacks externally.
\item Firewall nodes in EDoS Shield are susceptible to DDoS attack as it needs to handle all traffic of new IP addresses to verify the legitimacy of an incoming packet. In the case of a spoofed network traffic, this mechanism would suffer greatly from needing to verify each newly observed IP address. Meanwhile, EMM does not need to expose itself to the outside world, i.e., it works in a passive manner. EMM only need to interact to OFCs to update the flow tables after the detection of an ongoing attack.
\item EDoS Shield only provides protection for HTTP-based attacks whereas EMM provide protection to any type of general protocols attacks including HTTP, UDP and ICMP flooding attacks.
   
\end{itemize}

Moreover, EMM was also compared with some other existing EDoS mechanisms such as sPoW, and In-Cloud Scrubber based on their designed methodologies, resistance against EDoS (traffic types) and features that describes their functionalities. Both mechanisms were designed to mitigate the HTTP attacks. It orders to do so, sPoW performs packet filtering which is based on cryptography puzzles methodology whereas In-Cloud Scrubber does puzzle generation and verification process. However, these methods have certain limitations: 1) they require authentication from third party applications, 2) due to their design, delay is encountered while setting up the connection, 3) Likewise EDoS-Shield, these mechanisms can only mitigate HTTP based attacks. In contrast to that, EMM mechanism does not require an authentication process nor it relies on any kind of support from third party applications. Moreover, it is effective in terms of resource usage i.e. network, CPU, and memory.

In summary, Table \ref{tab:tab2} provides the comparison between the various state-of-the-art EDoS mitigation techniques, namely: EDoS Shield, sPoW, In-cloud Scrubber, and EMM.

\begin{table}[ht!]
\centering
    \caption{Comparison of various EDoS mitigation techniques.}
    \begin{adjustbox} {width=\textwidth}
    \begin{tabular}{c |>{\centering}m{3cm}| >{\centering}m{3cm}| >{\centering}m{2cm}| >{\centering}m{2cm}| c}
    \hline\hline 
    Techniques & Methodology & Mitigation Against & 3rd Party Auth. Requirement & Delay in Setting Up Connection  & Additional Overhead  \\
     \hline 
    EDoS-Sheild & Virtual firewall and Authentication & HTTP attacks & Yes	& Yes &	Yes \\
     \hline
     sPoW &	Packet filtering based on cryptography puzzles &HTTP attacks & Yes & Yes & Yes   \\  
    \hline
    In Cloud-Scrubber&	Puzzle generation and verification &HTTP attacks &Yes &	Yes &Yes \\
    \hline
    \textbf{EMM} & \textbf{Packet inspection} & \textbf{HTTP, UDP, ICMP attacks} &	\textbf{No} & \textbf{No}	& \textbf{No}\\
 \hline
    \end{tabular}
    \end{adjustbox}
    \label{tab:tab2}
\end{table}    

\section{Conclusion and Future Work}
\label{ssec:sec6}
In this study, the proposed technique is compared with existing EDoS detection and mitigation techniques by using various evaluation metrics, resistant to different type of EDoS attacks, and features that define their functionalities. The comparative results confirm that EMM can detect and mitigate EDoS attacks with higher accuracy rate than existing techniques. Unlike other state-of-the-art techniques, EMM detects an ongoing EDoS attack by conducting network monitoring. Due to its design, EMM requires very minimal overhead in deploying it in a cloud environment.

The evaluation was not only done with in comparison with state-of-the-art techniques, but most significantly, it was conducted in a real-world testbed environment that closely resembles a commercial cloud computing environment setup. EMM also fulfils its objectives of this study in coming up with an effective EDoS detection and mitigation mechanism.

Our area for future work would be to investigate and assert the feasibility of extending EMM to detect and mitigate EDoS/DDoS attacks as a SaaS model in cloud. Such a service could not only allow the users to ensure their bills are not inflated due to ongoing EDoS attacks, but also allow the cloud service providers to convince their new and existing customers that their cloud service incorporates EMM to provide a unique service that brings back the confidence to the users.


%\bibliographystyle{splncs04}
%\bibliography{bibliography.bib}

\begin{thebibliography}{10}
\providecommand{\url}[1]{\texttt{#1}}
\providecommand{\urlprefix}{URL }
\providecommand{\doi}[1]{https://doi.org/#1}

\bibitem{adamov2009truth}
Adamov, A., Erguvan, M.: The truth about cloud computing as new paradigm in it.
  In: 2009 International Conference on Application of Information and
  Communication Technologies. pp.~1--3. IEEE (2009)

\bibitem{baig2016controlled}
Baig, Z.A., Sait, S.M., Binbeshr, F.: Controlled access to cloud resources for
  mitigating economic denial of sustainability (edos) attacks  (2016)

\bibitem{bawa2017enhanced}
Bawa, P.S., Rehman, S.U., Manickam, S.: Enhanced mechanism to detect and
  mitigate economic denial of sustainability (edos) attack in cloud computing
  environments. INTERNATIONAL JOURNAL OF ADVANCED COMPUTER SCIENCE AND
  APPLICATIONS  \textbf{8}(9),  51--58 (2017)

\bibitem{bhardwaj2010cloud}
Bhardwaj, S., Jain, L., Jain, S.: Cloud computing: A study of infrastructure as
  a service (iaas). International Journal of engineering and information
  Technology  \textbf{2}(1),  60--63 (2010)

\bibitem{chapade2013securing}
Chapade, S., Pandey, K., Bhade, D.: Securing cloud servers against flooding
  based ddos attacks. In: 2013 International Conference on Communication
  Systems and Network Technologies. pp. 524--528. IEEE (2013)

\bibitem{chaudhary2018survey}
Chaudhary, D., Bhushan, K., Gupta, B.B.: Survey on ddos attacks and defense
  mechanisms in cloud and fog computing. International Journal of E-Services
  and Mobile Applications (IJESMA)  \textbf{10}(3),  61--83 (2018)

\bibitem{chowdhury2017edos}
Chowdhury, F.Z., Idris, M.Y.I., Kiah, M.L.M., Ahsan, M.M.: Edos eye: A game
  theoretic approach to mitigate economic denial of sustainability attack in
  cloud computing. In: 2017 IEEE 8th Control and System Graduate Research
  Colloquium (ICSGRC). pp. 164--169. IEEE (2017)

\bibitem{de2014using}
De~Oliveira, R.L.S., Schweitzer, C.M., Shinoda, A.A., Prete, L.R.: Using
  mininet for emulation and prototyping software-defined networks. In: 2014
  IEEE Colombian Conference on Communications and Computing (COLCOM). pp.~1--6.
  IEEE (2014)

\bibitem{green2011reconstructing}
Green, J., Juen, J., Fatemieh, O., Shankesi, R., Jin, D.K., Gunter, C.A.:
  Reconstructing hash reversal based proof of work schemes. In: LEET (2011)

\bibitem{hoff2008cloud}
Hoff, C.: Cloud computing security: From ddos (distributed denial of service)
  to edos (economic denial of sustainability). Rational Survivability  (2008)

\bibitem{hulboj2007packet}
Hulboj, M.M., Jurga, R.E.: Packet sampling and network monitoring (2007)

\bibitem{joshi2012securing}
Joshi, B., Vijayan, A.S., Joshi, B.K.: Securing cloud computing environment
  against ddos attacks. In: 2012 International Conference on Computer
  Communication and Informatics. pp.~1--5. IEEE (2012)

\bibitem{khor2009spow}
Khor, S.H., Nakao, A.: spow: On-demand cloud-based eddos mitigation mechanism.
  In: HotDep (Fifth Workshop on Hot Topics in System Dependability) (2009)

\bibitem{kumar2012mitigating}
Kumar, M.N., Sujatha, P., Kalva, V., Nagori, R., Katukojwala, A.K., Kumar, M.:
  Mitigating economic denial of sustainability (edos) in cloud computing using
  in-cloud scrubber service. In: 2012 Fourth International Conference on
  Computational Intelligence and Communication Networks. pp. 535--539. IEEE
  (2012)

\bibitem{kuyoro2011cloud}
Kuyoro, S., Ibikunle, F., Awodele, O.: Cloud computing security issues and
  challenges. International Journal of Computer Networks (IJCN)  \textbf{3}(5),
   247--255 (2011)

\bibitem{masood2013edos}
Masood, M., Anwar, Z., Raza, S.A., Hur, M.A.: Edos armor: a cost effective
  economic denial of sustainability attack mitigation framework for e-commerce
  applications in cloud environments. In: INMIC. pp. 37--42. IEEE (2013)

\bibitem{popovic2010cloud}
Popovi{\'c}, K., Hocenski, {\v{Z}}.: Cloud computing security issues and
  challenges. In: The 33rd International Convention MIPRO. pp. 344--349. IEEE
  (2010)

\bibitem{sengar2008detecting}
Sengar, H., Wang, H., Wijesekera, D., Jajodia, S.: Detecting voip floods using
  the hellinger distance. IEEE transactions on parallel and distributed systems
   \textbf{19}(6),  794--805 (2008)

\bibitem{shalimov2013advanced}
Shalimov, A., Zuikov, D., Zimarina, D., Pashkov, V., Smeliansky, R.: Advanced
  study of sdn/openflow controllers. In: Proceedings of the 9th central \&
  eastern european software engineering conference in russia. p.~1. ACM (2013)

\bibitem{shannon1948note}
Shannon, C.E.: A note on the concept of entropy. Bell System Tech. J
  \textbf{27}(3),  379--423 (1948)

\bibitem{shawahna2018edos}
Shawahna, A., Abu-Amara, M., Mahmoud, A., Osais, Y.E.: Edos-ads: An enhanced
  mitigation technique against economic denial of sustainability (edos)
  attacks. IEEE Transactions on Cloud Computing  (2018)

\bibitem{singh2014survey}
Singh, P., Manickam, S., Rehman, S.U.: A survey of mitigation techniques
  against economic denial of sustainability (edos) attack on cloud computing
  architecture. In: Proceedings of 3rd International Conference on Reliability,
  Infocom Technologies and Optimization. pp.~1--4. IEEE (2014)

\bibitem{sqalli2011edos}
Sqalli, M.H., Al-Haidari, F., Salah, K.: Edos-shield-a two-steps mitigation
  technique against edos attacks in cloud computing. In: 2011 Fourth IEEE
  International Conference on Utility and Cloud Computing. pp. 49--56. IEEE
  (2011)

\bibitem{swami2019software}
Swami, R., Dave, M., Ranga, V.: Software-defined networking-based ddos defense
  mechanisms. ACM Computing Surveys (CSUR)  \textbf{52}(2), ~28 (2019)

\bibitem{velte2010cloud}
Velte, A.T., Velte, T.J., Elsenpeter, R.C., Elsenpeter, R.C.: Cloud computing:
  a practical approach. McGraw-Hill New York (2010)

\bibitem{zissis2012addressing}
Zissis, D., Lekkas, D.: Addressing cloud computing security issues. Future
  Generation computer systems  \textbf{28}(3),  583--592 (2012)

\end{thebibliography}

\end{document}